\documentclass{article}
\usepackage{spconf,amsmath,amssymb,graphicx}
\usepackage{booktabs}
\usepackage{multirow}
\usepackage{hyperref}

\title{Sori: Streamable and Controllable Model for Non-musical Audio to Symbolic Music Transformation}

\name{Kyungsu Kim \quad Yejin Kim \quad Kyogu Lee}
\address{Music and Audio Research Group (MARG), Seoul National University}
\begin{document}
\maketitle

\begin{abstract}
We introduce \textit{Non-musical Audio to Symbolic Music Transformation} (NASMT), a novel task that converts non-musical audio into symbolic music representations.
Unlike automatic music transcription, which assumes musical input and a unique correct score, NASMT is inherently one-to-many and requires evaluation along two axes: input correspondence and musicality.
We present Sori, an encoder--decoder Transformer that addresses this task through domain adversarial training (DAT), which bridges the gap between musical and non-musical audio in the encoder's feature space, and classifier-free guidance (CFG) with a continuous parameter~$\gamma$ that controls the trade-off between correspondence and musicality at inference time.
Across objective metrics and a listening study, Sori provides effective continuous control via $\gamma$ and compares favorably with instrument-agnostic transcription models.
Sori supports real-time streaming on a consumer CPU.
Audio examples are available at \url{https://soridemo0320.github.io/sori-demo/}.
\end{abstract}


\section{Introduction}
\label{sec:intro}

The musicality of a sound depends not on its acoustic properties but on how the
listener interprets it~\cite{reybrouck2014music}.
This idea has motivated artistic practices that reinterpret everyday sounds as musical material, from musique concr\`ete~\cite{schaeffer1966traite} to contemporary sound-based composition.
Motivated by this rich tradition of finding musicality in everyday sound, we propose \textit{Non-musical Audio to Symbolic Music Transformation} (NASMT), a task that converts non-musical audio into symbolic music representations.

\begin{figure}[t]
  \centering
  \includegraphics[width=\linewidth]{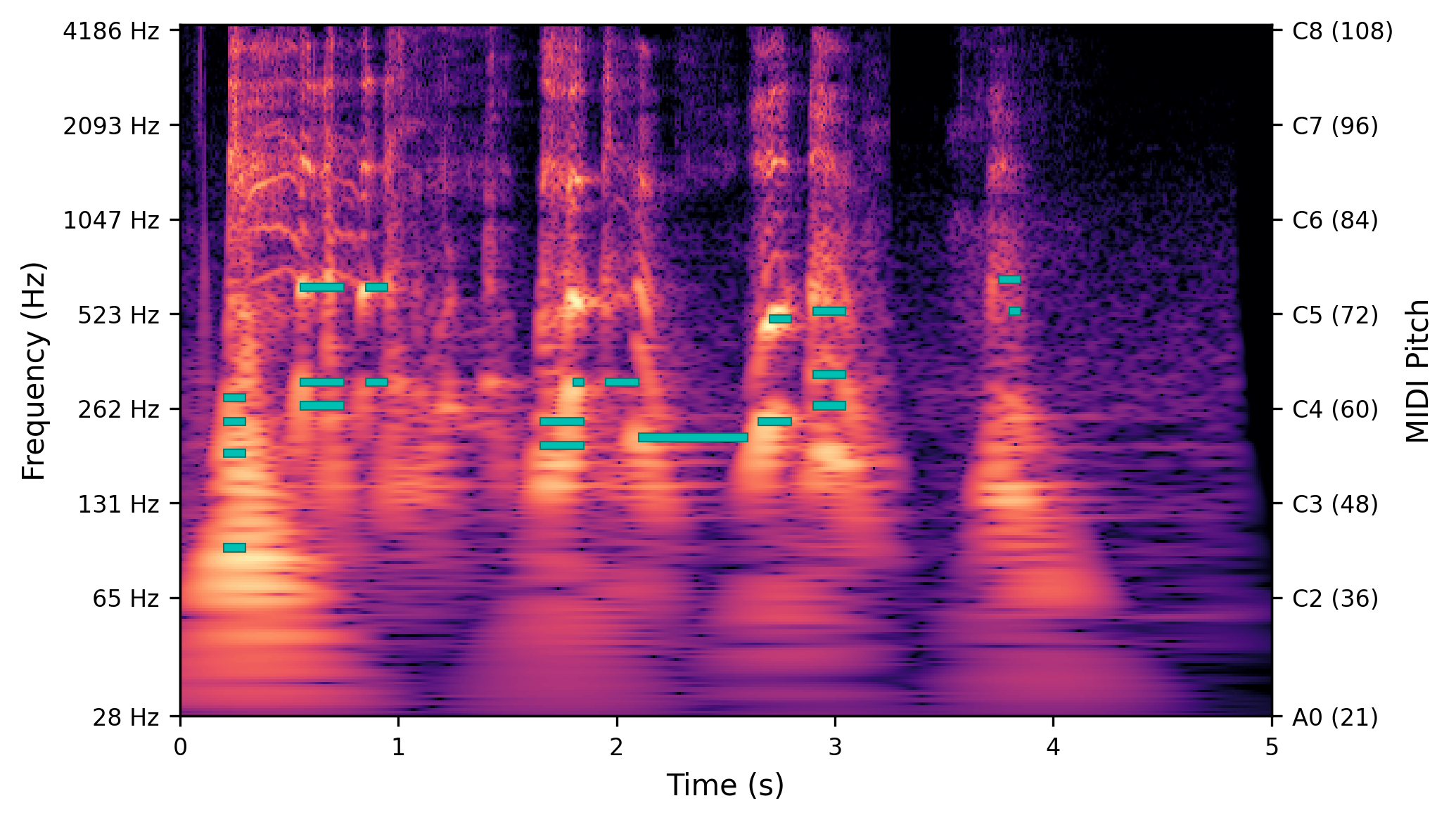}
  \caption{Illustration of the NASMT task. A CQT spectrogram of non-musical input (a cough sound) is overlaid with the MIDI piano roll generated by Sori. The output aligns with the temporal and harmonic structure of the input while forming musically coherent note patterns. Note that the CQT display is used for visualization only; the model operates on log-Mel spectrograms.}
  \label{fig:example}
\end{figure}

Unlike automatic music transcription (AMT)~\cite{benetos2019automatic}, which assumes a one-to-one mapping between audio and score, NASMT is inherently one-to-many: given a recording of rain or footsteps, many musically valid interpretations exist.
Figure~\ref{fig:example} illustrates the task: a non-musical input (a cough sound) is transformed into a MIDI sequence that aligns with the input's temporal and harmonic structure while forming musically coherent note patterns.
We evaluate NASMT outputs along two axes: \textit{input correspondence}, the degree to which the output reflects the spectral and temporal characteristics of the input, and \textit{musicality}, the coherence and plausibility of the resulting music.
These two objectives stand in fundamental tension---faithfully preserving input structure tends to reduce musical coherence, while prioritizing musicality requires departing from the source signal.

To tackle this task, we present \textit{Sori} (meaning ``sound'' in Korean), an encoder-decoder Transformer for streamable audio-to-MIDI transformation.
Since non-musical audio lacks paired MIDI annotations, we train the decoder exclusively on musical audio and apply domain adversarial training~\cite{ganin2016domain, sato2024annotation} so that the encoder maps non-musical inputs close to musical audio in the feature space, allowing the decoder to produce musically coherent output from non-musical input audio.
To control the inherent trade-off between input correspondence and musicality, we adopt classifier-free guidance (CFG)~\cite{ho2022classifier, sanchez2023stay} with a continuous control parameter~$\gamma$, and introduce a nested autoregressive decoder that ensures CFG operates on the joint event distribution.
The model is lightweight (153M parameters at inference) and supports real-time inference with 64~ms algorithmic latency---making it suitable for live performance and sound installation contexts.

Our contributions are:
\textbf{(1)}~We propose the NASMT task and a two-axis evaluation framework that explicitly recognizes its one-to-many nature.
\textbf{(2)}~We adapt DAT to bridge musical and non-musical audio domains, enabling musically coherent output from non-musical input audio.
\textbf{(3)}~We enable continuous inference-time control of the correspondence--musicality trade-off through classifier-free guidance.

\section{Method}\label{sec:method}
\begin{figure}[t]
  \centering
  \includegraphics[width=\linewidth]{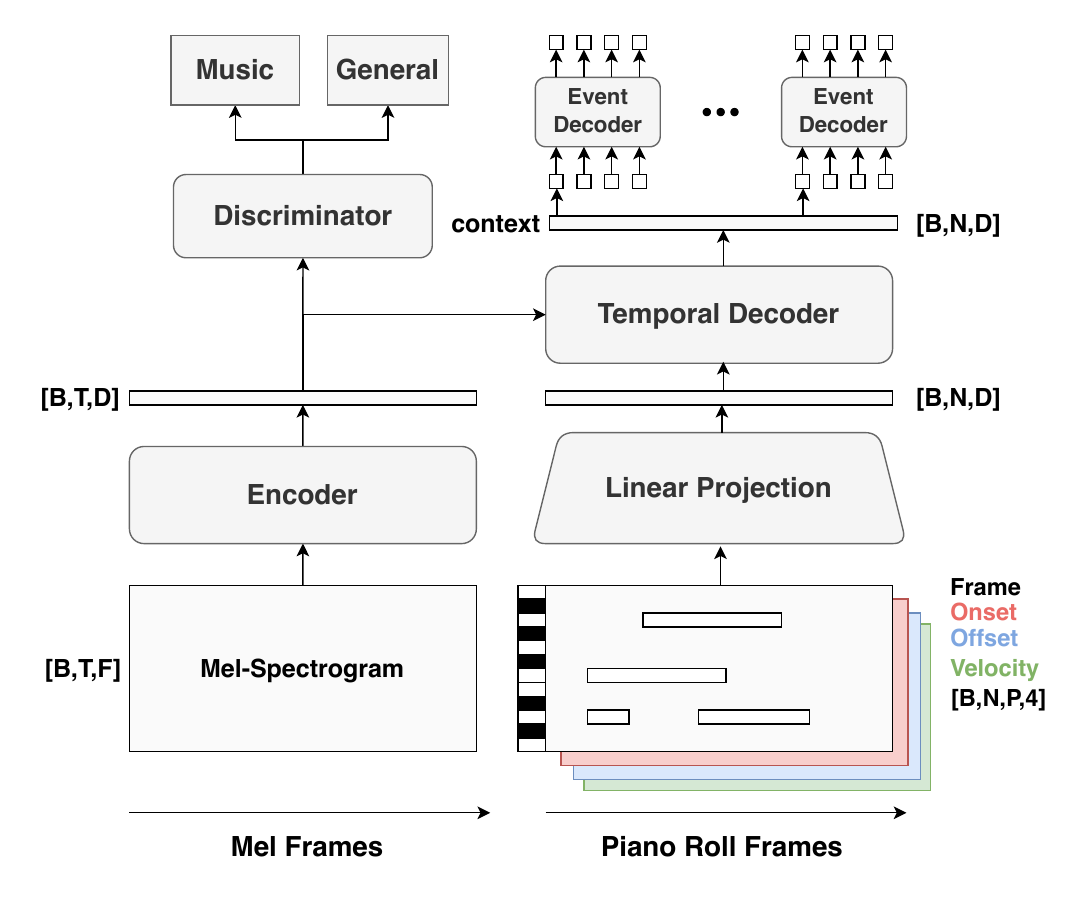}
  \caption{Sori architecture. The encoder maps log-Mel frames to latent features; the Temporal Decoder cross-attends over encoder output to produce per-timestep context vectors; the Event Decoder generates note events autoregressively within each timestep.}
  \label{fig:architecture}
\end{figure}

\subsection{Task Formulation}\label{subsec:task}
We define \textit{Non-musical Audio to Symbolic Music Transformation} (NASMT) as the task of transforming a non-musical audio signal $a \in \mathcal{A}$, such as speech, environmental noise, or everyday sounds, into a symbolic music representation $m \in \mathcal{M}$.
Unlike AMT, which assumes a one-to-one mapping from musical audio to a unique correct score, NASMT is inherently one-to-many: many musically valid symbolic interpretations may plausibly correspond to the same non-musical input.

We therefore evaluate NASMT outputs along two complementary axes.
\textit{Input correspondence}~$C(m, a) \in \mathbb{R}$ quantifies how faithfully the symbolic output reflects the temporal and spectral structure of the input audio.
\textit{Musicality}~$M(m) \in \mathbb{R}$ quantifies the musical coherence and plausibility of the output, independent of the input.
These two objectives are in fundamental tension: closely following the input signal can reduce musical coherence, while improving musicality can weaken correspondence to the source.

We formalize NASMT as the optimization:
\begin{equation}\label{eq:task}
\hat{m} = \arg\max_{m \in \mathcal{M}} \bigl[\lambda \, C(m, a) + (1 - \lambda) \, M(m)\bigr],
\end{equation}
Here $C$ and $M$ are inherently subjective and cannot be defined in closed form.
In practice, we approximate them through objective proxy metrics and a listening study (Section~\ref{ssec:setup}).
$\lambda \in [0,1]$ governs the correspondence--musicality trade-off.
Rather than fixing this trade-off at training time, Sori exposes it as a continuous inference-time control via the CFG scale $\gamma$ (Section~\ref{ssec:training}), allowing a single model to operate anywhere along the trade-off axis.

\subsection{Model Architecture}
\label{ssec:arch}

Sori consists of an encoder and two decoders (Fig.~\ref{fig:architecture}).
The 16~kHz input audio is converted to 512~bin log-Mel spectrograms ($n_\text{fft}=2048$, hop $=800$, corresponding to a 50~ms hop), which are processed by an 8-layer pre-norm Transformer encoder ($d=768$, 8 heads).
The \textit{Temporal Decoder} (8 layers, $d=768$, 8 heads) uses causal cross-attention to produce a context vector $\mathbf{c}_t$ for each encoder timestep.
The \textit{Event Decoder} (4 layers, $d=512$, 8 heads) then autoregressively generates a variable number of note events at each timestep.

Unlike dense tokenization~\cite{gardner2022mt3}, which represents music as a sequence of densely ordered tokens, we adopt a frame-aligned sparse representation~\cite{wei2025streaming} where events are decoded per timestep, enabling frame-by-frame streaming.
The vocabulary contains 530 tokens, consisting of 88 note-off tokens, 440 note-on tokens covering 88 pitches and 5 quantized velocity bins, and one token each for EOS and PAD.
The events decoded at each timestep are converted into a 4-channel piano roll $\mathbf{r}_t \in \mathbb{R}^{P \times 4}$, which is fed to the Temporal Decoder at the next timestep.
The four channels represent onset, offset, velocity, and frame activity.
For each pitch, unoccupied entries are set to $-1$; for active notes, the onset and offset channels are set to $1$, and the velocity channel stores a value in $[0,1]$ corresponding to the quantized note velocity.
The frame-activity channel is set to $1$ for all currently active pitches, explicitly indicating ongoing notes and providing temporal memory analogous to the active onset set of Wei et al.~\cite{wei2025streaming}.

To model the joint distribution of note events within each timestep, we adopt a nested autoregressive framework~\cite{kim2024nestedmusictransformer}.
Each $e_t^k$ denotes a note-on, note-off, or EOS event:
\begin{equation}\label{eq:temporal}
P(e_{1:T} \mid a)
= \prod_{t=1}^{T} P(e_t^1, \ldots, e_t^K \mid e_{<t}, a_{\leq t}).
\end{equation}
Each per-timestep joint is factorized autoregressively:
\begin{equation}\label{eq:event}
P(e_t^1, \ldots, e_t^K \mid e_{<t}, a_{\leq t})
= \prod_{k=1}^{K} P(e_t^k \mid e_t^{<k}, \mathbf{c}_t),
\end{equation}
where $e_t^{<k}$ denotes preceding events within timestep $t$ and $\mathbf{c}_t$ summarizes the past event and audio context.
This factorization allows CFG to be applied at each within-timestep event-decoding step while retaining a bounded per-frame decoding budget (Sec.~\ref{ssec:training}).

To bound the per-timestep computation for streaming inference, we set $K=4$.
If more than four events occur at a timestep, the surplus events are deferred to the next frame via a binary spillover flag.
Thus, each timestep requires at most four event-decoding steps, while 99.52\% of training timesteps contain four or fewer events and require no spillover.
Together with ALiBi positional encoding~\cite{press2022alibi} and an attention-sink KV cache~\cite{xiao2024streaming}, this enables bounded-latency streaming beyond the training context length.

\subsection{Training and Generation}
\label{ssec:training}

\noindent \textbf{Domain Adversarial Training.} \enspace
MIDI annotations exist only for musical audio, so the decoder is trained exclusively on musical data.
At inference time, non-musical input introduces a domain shift that can degrade output musicality.
We apply domain adversarial training (DAT)~\cite{ganin2016domain} to the encoder, following the setting of Sato and Akama~\cite{sato2024annotation} to align musical and non-musical audio representations.
The discriminator predicts whether encoder features originate from musical or non-musical audio using a learned CLS representation.
Training alternates between discriminator updates with binary domain labels and encoder/decoder updates that optimize transcription while encouraging domain-confused encoder representations.

\noindent \textbf{Classifier-Free Guidance.} \enspace
Classifier-free guidance (CFG)~\cite{ho2022classifier,sanchez2023stay} forms guided logits by affinely combining conditional and unconditional logits at each decoding step:
\begin{equation}\label{eq:cfg}
\tilde{\boldsymbol{\ell}} = \boldsymbol{\ell}_{\text{uncond}} + \gamma \cdot \left(\boldsymbol{\ell}_{\text{cond}} - \boldsymbol{\ell}_{\text{uncond}}\right).
\end{equation}
At $\gamma=0$, generation uses the unconditional branch and is independent of the input; at $\gamma=1$, it performs standard conditional generation; and at $\gamma>1$, the influence of the input is amplified.
During training, we drop the input conditioning with probability $p_\text{drop}\!=\!0.2$, replacing the encoder output with a learned null embedding to train the unconditional branch required by CFG.

Applied to our task, CFG provides continuous control over the correspondence--musicality trade-off, ranging from unconditional generation at $\gamma=0$ to increasingly strong input conditioning as $\gamma$ increases.

\begin{table*}[t]
\centering
\small
\setlength{\tabcolsep}{4.2pt}
\begin{tabular}{@{}ll ccccc ccccc ccc@{}}
\toprule
& & \multicolumn{5}{c}{\textbf{Sori}}
& \multicolumn{5}{c}{\textbf{Sori (w/o DAT)}}
& \multicolumn{3}{c}{\textit{Reference}} \\
\cmidrule(lr){3-7} \cmidrule(lr){8-12} \cmidrule(l){13-15}
& $\gamma$ & 0.0 & 0.5 & 1.0 & 2.0 & 3.0
& 0.0 & 0.5 & 1.0 & 2.0 & 3.0
& Basic Pitch & Deep Salience & TimbreTrap \\
\midrule
\multirow{3}{*}{\textit{Objective}}
& OnF1,$\uparrow$
& 0.087 & 0.236 & 0.327 & 0.363 & 0.368
& 0.282 & 0.340 & 0.381 & \textbf{0.390} & 0.386
& 0.210 & 0.300 & 0.167 \\
& MSS,$\downarrow$
& 57.2 & 32.7 & 18.3 & 13.1 & 11.8
& 25.4 & 17.1 & 11.3 & \textbf{10.1} & 10.2
& 32.8 & 21.7 & 33.1 \\
& FMD,$\downarrow$
& 143.6 & \textbf{33.6} & 51.3 & 161.8 & 248.0
& 96.7 & 96.4 & 191.3 & 288.7 & 324.0
& 181.8 & 195.2 & 472.7 \\
\addlinespace[3pt]
\multirow{2}{*}{\textit{Subjective}}
& Cor.,$\uparrow$
& 1.20 & 1.64 & 2.61 & \textbf{2.97} & 2.52
& 1.39 & 1.79 & 2.50 & 2.66 & 2.54
& 2.35 & 2.63 & 2.51 \\
& Mus.,$\uparrow$
& 2.19 & \textbf{3.34} & 3.27 & 2.59 & 2.20
& 2.78 & 3.32 & 2.72 & 2.07 & 1.98
& 2.15 & 2.37 & 1.51 \\
\bottomrule
\end{tabular}
\caption{Objective and subjective evaluation on the AudioSet evaluation set. OnF1: onset F1 averaged over 11 DSP-based onset detectors with a $\pm100$~ms tolerance. MSS: Multi-Scale Spectrogram distance (lower = more similar). FMD: Fr\'echet Music Distance via CLaMP\,2. Cor./Mus.: mean listener ratings of input correspondence and musicality. Instrument-agnostic single-track AMT models are used as reference methods.}
\label{tab:results}
\end{table*}

\section{Experiments}

\label{sec:exp}

\subsection{Experimental Setup}
\label{ssec:setup}

\noindent \textbf{Datasets.}\enspace
We construct a musical dataset following the Nlakh pipeline~\cite{kim2023nlakh}, using sample-based instruments derived from NSynth~\cite{engel2017nsynth} to render Lakh MIDI files.
We generate 10k segments per instrument, yielding 9.53M five-second training segments across 953 instruments and 530k held-out test segments across 53 instruments at 16kHz.
Instruments and MIDI files are disjoint across splits, with the instrument split following the train/test partition of NSynth.
The test set is used as the reference distribution for FMD computation.
For DAT, we use the downloadable portion of AudioSet's~\cite{gemmeke2017audioset} unbalanced training split (1.65M five-second clips) for adversarial training and its evaluation split (17k clips) for evaluation.

\noindent \textbf{Implementation Details.}\enspace
We train Sori for 300k iterations with batches of 16 musical and 16 non-musical examples.
We use AdamW with a weight decay of 0.1 and learning rates of $3 \times 10^{-5}$ for the encoder and decoder and $3 \times 10^{-6}$ for the discriminator.
A 10k-iteration linear warmup is followed by learning-rate decays at 30k and 60k iterations.
For the first 50k iterations, the encoder and decoder are trained using only the transcription objective.
Thereafter, we alternate five discriminator updates with one encoder--decoder update, weighting the adversarial loss by 0.01.
The \textit{Sori (w/o DAT)} ablation omits adversarial training.

\noindent \textbf{Metrics.}\enspace
\textit{Onset F1} (OnF1) measures temporal correspondence between the input audio and generated MIDI.
For each of 11 DSP-based onset detection algorithms~\cite{bock2016madmom}, we detect onsets in the input audio, treat them as reference events, and compute F1 against note onset times from the generated MIDI with a $\pm100$~ms tolerance; the resulting scores are averaged across detectors.
We use only DSP-based onset detectors to avoid bias from neural detectors trained on musical audio.

\textit{Multi-Scale Spectrogram similarity} (MSS)~\cite{engel2020ddsp} renders the
output MIDI with FluidSynth (program 0, FluidR3 GM soundfont) and computes
log-magnitude L1 distance against the input audio across five spectral scales; lower MSS indicates greater spectral correspondence.
We verified that relative rankings among models are mostly consistent across most General MIDI instrument programs.

\textit{Fr\'echet Music Distance} (FMD)~\cite{retkowski2024fmd} computes the Fr\'echet distance between CLaMP\,2~\cite{wu2024clamp2} embeddings of generated MIDI and the held-out Nlakh test set.
FMD therefore measures proximity to the musical reference distribution rather than reconstruction fidelity.

We conducted an IRB-approved listening study (IRB No.~2512/003-007) via Prolific~\cite{palan2018prolific} using a 5-point Likert scale.
Each participant evaluated 20 sets of 13 model conditions per axis, totaling 520 trials, with stereo input--output pairs for correspondence and monaural outputs for musicality.
As a task-specific attention check, we excluded 12 participants whose mean correspondence rating for the unconditional $\gamma=0$ condition exceeded 2.0, retaining $N=21$.

\noindent \textbf{Reference models.}\enspace
Since no prior systems directly address this task, we use three instrument-agnostic single-track AMT models as reference methods: Basic Pitch~\cite{bittner2022lightweight}, Deep Salience~\cite{bittner2017deep}, and TimbreTrap~\cite{cwitkowitz2024timbretrap}.
As these models are designed for musical input, their results serve as reference points rather than direct comparisons.

\subsection{Results}
Table~\ref{tab:results} summarizes the objective and subjective evaluation results on the AudioSet evaluation set.
As $\gamma$ increases, objective input correspondence generally improves in both temporal and spectral dimensions, as reflected by higher OnF1 and lower MSS, while subjective correspondence follows the same trend up to moderate-to-strong guidance.
In contrast, musicality is strongest under moderate guidance and decreases as conditioning becomes stronger.
At $\gamma=3.0$, objective correspondence shows diminishing gains and subjective correspondence also declines, indicating diminishing returns from excessive guidance.
The effect of $\gamma$ is significant for both correspondence and musicality for Sori and Sori (w/o DAT) (Friedman test, $p<0.001$ for all).

Compared with the instrument-agnostic AMT reference methods, Sori at $\gamma=0.5$ achieves the best FMD and the highest subjective musicality among all evaluated conditions.
At $\gamma=1.0$, Sori outperforms all three reference methods on both objective correspondence metrics (OnF1 and MSS) as well as FMD, while achieving comparable subjective correspondence and higher perceived musicality.
This suggests that moderate conditioning allows Sori to preserve both temporal and spectral input structure while producing outputs that remain close to the musical reference distribution.

The DAT ablation reveals a consistent trade-off between input correspondence and musicality.
Removing DAT generally improves both temporal and spectral correspondence, as reflected by higher OnF1 and lower MSS, but degrades FMD and subjective musicality under stronger guidance.
These results indicate that DAT shifts the correspondence--musicality trade-off toward musicality as input conditioning becomes stronger, rather than uniformly improving performance across all operating points.

Overall, $\gamma$ provides a practical continuous control rather than requiring a fixed operating point.
Moderate guidance offers the most useful trade-off between correspondence and musicality, while stronger guidance progressively prioritizes input tracking.

For streaming performance, we define the real-time factor (RTF) as processing time divided by audio duration.
Sori achieves a mean RTF of 0.82 (maximum 0.90) on an Apple M3 CPU at 50~ms frame resolution.
The 64~ms algorithmic latency is dominated by the centered STFT, whose $n_{\text{fft}}/2=1024$-sample lookahead corresponds to 64~ms at 16~kHz.
The nested decoder introduces no additional algorithmic lookahead because all $K=4$ event-decoding steps are completed within each frame budget.
Together, ALiBi positional encoding and the attention-sink KV cache enable continuous real-time inference beyond the training context length.
\section{Conclusion \& Limitations}
\label{sec:conclusion}
We introduced Non-musical Audio to Symbolic Music Transformation (NASMT) and presented Sori, a controllable and streamable model for transforming non-musical audio into symbolic music.
Classifier-free guidance provides continuous inference-time control over the correspondence--musicality trade-off, while domain adversarial training helps preserve musical coherence under stronger input conditioning.
The nested autoregressive decoder enables bounded per-frame computation, allowing Sori to operate in real time on a consumer CPU.
Several limitations remain: the AudioSet splits are used without filtering music-related clips, so the evaluation set is not exclusively non-musical; the contribution of the nested autoregressive design is not separately ablated; and FMD is measured against Nlakh, which is derived from the same underlying Lakh MIDI corpus used for training and may therefore not fully reflect musicality beyond that distribution.
Future work should address these limitations and explore broader forms of musical control beyond the correspondence--musicality trade-off.
\bibliographystyle{IEEEbib}
\bibliography{refs}

\end{document}